\documentclass[%
reprint,
superscriptaddress,
 amsmath,amssymb,nofootinbib,
 aps,
 twocols,
]{revtex4-1}

\usepackage[title]{appendix}
\usepackage{verbatim}
\usepackage{graphicx}% Include figure files
\usepackage{dcolumn}% Align table columns on decimal point
\usepackage{bm}
\usepackage{mathtools}
\usepackage{float}
\makeatletter
\let\newfloat\newfloat@ltx
\makeatother
\usepackage{algorithm} 
\usepackage{algpseudocode}  
\usepackage{float}                                    %%%
\usepackage{graphicx} 
\usepackage[font=small,labelfont=bf,justification=raggedright,format=plain]{caption}
\usepackage{hyperref}                                 %%%
\usepackage[usenames, dvipsnames]{color}
\usepackage{color}
\usepackage{listings}
\usepackage{enumerate}
\usepackage{multirow}

\definecolor{dkgreen}{rgb}{0,0.6,0}
\definecolor{gray}{rgb}{0.5,0.5,0.5}
\definecolor{mauve}{rgb}{0.58,0,0.82}

\begin{document}

%\preprint{APS/123-QED}

\title{Fast, accurate, and system-specific variable-resolution modelling of proteins}

%breaks with \\

%\author{Aaa Bbb}
\author{Raffaele Fiorentini}
\affiliation{Physics Department, University of Trento, via Sommarive, 14 I-38123 Trento, Italy}
\affiliation{INFN-TIFPA, Trento Institute for Fundamental Physics and Applications, I-38123 Trento, Italy}
\author{Thomas Tarenzi}%
\affiliation{Physics Department, University of Trento, via Sommarive, 14 I-38123 Trento, Italy}
\affiliation{INFN-TIFPA, Trento Institute for Fundamental Physics and Applications, I-38123 Trento, Italy}
\author{Raffaello Potestio}%
 \email{raffaello.potestio@unitn.it}
\affiliation{Physics Department, University of Trento, via Sommarive, 14 I-38123 Trento, Italy}
\affiliation{INFN-TIFPA, Trento Institute for Fundamental Physics and Applications, I-38123 Trento, Italy}

\date{\today}% It is always \today, today,

\begin{abstract}
In recent years, a few multiple-resolution modelling strategies have been proposed, in which functionally relevant parts of a biomolecule are described with atomistic resolution, while the remainder of the system is concurrently treated using a coarse-grained model. In most cases, the parametrisation of the latter requires lengthy reference all-atom simulations and/or the usage of off-shelf coarse-grained force fields, whose interactions have to be refined to fit the specific system under examination. Here, we overcome these limitations through a novel multi-resolution modelling scheme for proteins, dubbed coarse-grained anisotropic network model for variable resolution simulations, or CANVAS. This scheme enables the user-defined modulation of the resolution level throughout the system structure; a fast parametrisation of the potential without the necessity of reference simulations; and the straightforward usage of the model on the most commonly used molecular dynamics platforms. The method is presented and validated on two case studies, the enzyme adenylate kinase and the therapeutic antibody pembrolizumab, by comparing results obtained with the CANVAS model against fully atomistic simulations. The modelling software, implemented in python, is made freely available for the community on a collaborative github repository.
\end{abstract}

\maketitle

\begin{figure}[t]
\centering
\includegraphics[width=\columnwidth]{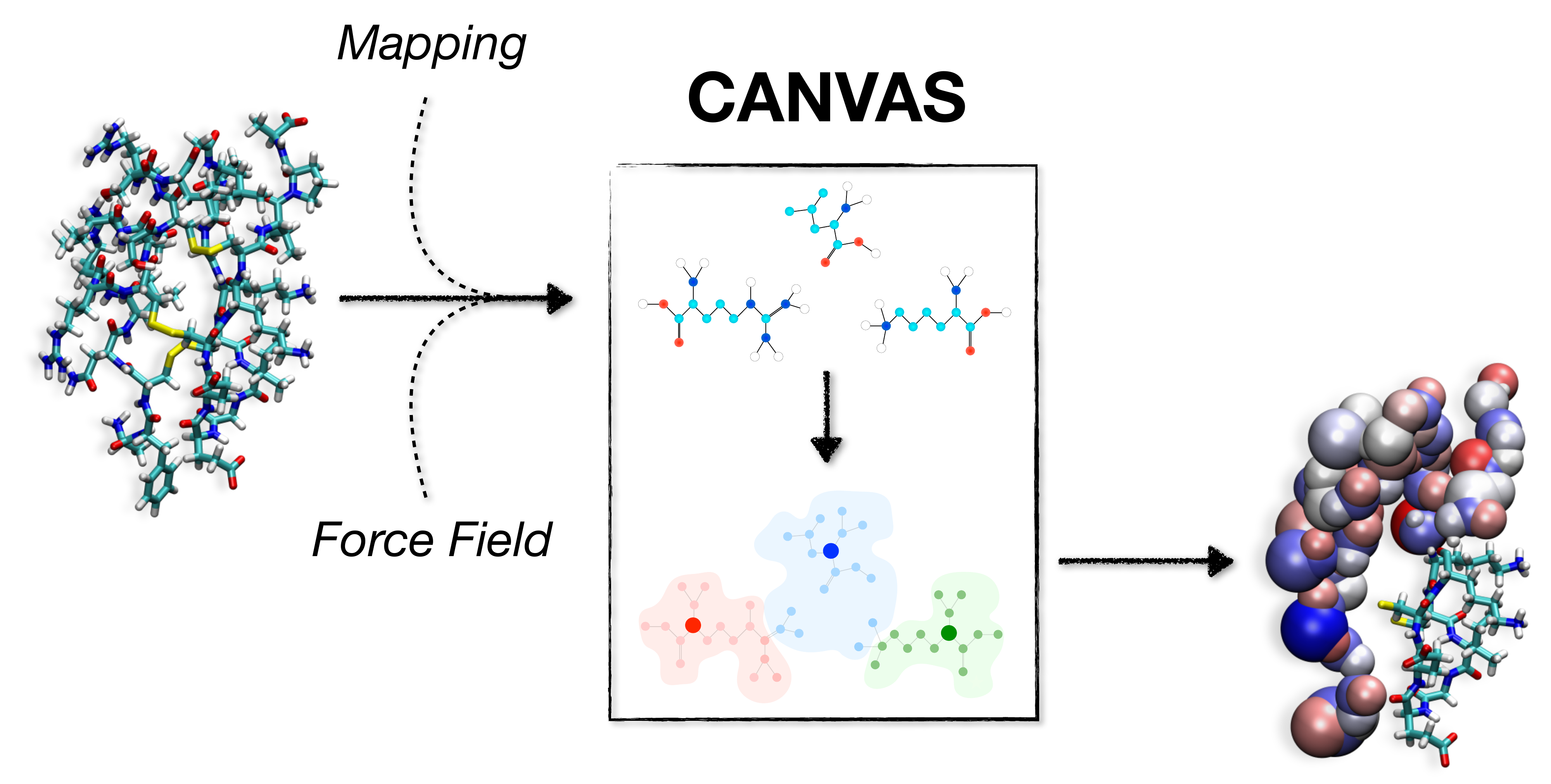}
\caption*{\footnotesize Table-of-contents entry.}
\label{fig:toc}
\end{figure}

\section{Introduction}\label{sect:intro}

Steady improvements in high performance computing hardware and molecular dynamics (MD) simulation software over several decades have ushered impressive advancements in the computer-aided investigation of soft and biological matter systems, in particular macromolecules of biological origin such as lipids, proteins, and nucleic acids \cite{stone2010gpu, lazim2020advances, shaw2021anton}. At the same time, a detailed modelling of molecular systems, in which each atom is described as an interaction center, often turns out to be inconvenient or even undesirable, on the one hand due to the major computing and data storage requirements, on the other because of the effort in analysing the simulation outcome. To overcome both limitations, simplified, {\it coarse-grained} (CG) models \cite{B912027K,Kamerlin,TAKADA2012,C2CP40934H,Noid-2013-perp,giulini2021system} are frequently employed, in which several atoms are lumped in effective interaction sites. CG models enable the simulation of larger systems over longer time scales, thanks to a smoother (free) energy profile and fewer degrees of freedom with respect to all-atom representations.

Coarse-grained models have been successfully employed for a number of biologically and pharmacologically relevant applications.
These include the study of spontaneous protein-ligand binding~\cite{souza2020protein}, where both the macromolecule, the ligand, and the solvent are modelled in a coarse-grained fashion. The approach proved useful for the identification of binding pockets and the estimation of binding free energies on a number of systems; however, the simplified representation of the ligand requires a not-so-obvious new parameterization of the interactions, and limits the distinction between similar molecules (as in the case of enantiomers)~\cite{souza2020protein}. In addition, the employment of a coarse-grained solvent model limits the accuracy in the case single water molecules are actively involved in the stabilization of the ligand in the binding site. Similarly, a number of coarse-grained force-fields, including, among the others, the MARTINI \cite{marrink2019computational, marrink2022two}, SIRAH \cite{machado2019sirah, patel2018fast}, AWSEM \cite{davtyan2012awsem} and Scorpion \cite{basdevant2007coarse} force fields, have been used to investigate protein-protein interactions, providing accurate results in terms of binding free energies. However, those models prevent an even coarser representation of the protein interfaces, which might be desirable in the case of very large protein assemblies; furthermore, they do not provide an accurate description of the system if specific atomistic details, possibly crucial for the properties or behaviour of interest, are effectively integrated out in the low-resolution model.

Hence, while all-atom models provide the necessary accuracy at the expenses of substantial computational resources, CG models enable a more efficient and intelligible representation of the system at the cost of losing possibly crucial detail. Although several problems in computational biophysics can be tackled with one of these two methods, many open questions remain that necessitate an approach at the interface between chemical accuracy and computational efficiency. In this regard, methods have been developed in which molecules described at different resolution are simultaneously simulated within the same setup. Examples include the coupling of  MARTINI with atomistic force fields \cite{rzepiela2011hybrid, wassenaar2013mixing}; the simulation of atomistic proteins and nucleic acids in multi-resolution solvent with the SWINGER algorithm \cite{zavadlav2018open, zavadlav2018multiscale}; and the simulation of soluble proteins with the PACE force field, which has been developed with the specific aim of coupling united-atom protein models with a coarse-grained solvent representation \cite{han2010pace, han2012further}. Pushing the ``resolution mix'' even further, in some applications it might be desirable to couple different levels of detail within the same biomolecule, in order to limit the computationally expensive high-resolution modelling to a subset of protein residues or nucleic acid base-pairs. This approach was pioneered by the quantum mechanics/molecular mechanics (QM/MM) methods \cite{QM-MM-2, QM-MM-3, QM-MM-4, cui2021biomolecular, vennelakanti2022harder}, which allow a connection between a small region where {\it ab initio} models are used, and a classical all-atom description in the remainder of the system. Along the same lines, several methodologies have been developed to couple atomistic and coarse-grained levels of resolution within the same simulation set-up, and even within the same molecular structure. For example, in the Molecular Mechanics/Coarse-Grained (MM/CG) scheme developed in 2005 by Neri {\it et.al} \cite{neri2005coarse}, the atomistically detailed active site is incorporated into a coarse-grained G{\=o}-like model, which aims at reproducing the correct conformational fluctuations of the full protein \cite{schneider2018predicting}. The MM/CG method was later tailored for the simulation of membrane protein/ligand complexes \cite{leguebe2012hybrid}, and in the last version of the method, dubbed open-boundary MM/CG \cite{tarenzi2019open}, the dual-resolution description of the protein is coupled with an adaptive multiscale model of the solvent, namely the Hamiltonian adaptive resolution scheme (H-AdResS) \cite{Donadio_PRL_2013-hadres_molliq, tarenzi2017open}; in the latter, regions of different resolution are defined in the simulation box, allowing water molecules to change their resolution \textit{on the fly} when diffusing from one region to the other. More recently, a similar method \cite{Kremer_Proteins_2016-lys_multires, fiorentini2020ligand} employed a high-resolution force field in small regions of a protein, most notably the active site, while treating the remainder in a coarse-grained fashion, e.g. as an elastic network model (ENM) \cite{Tirion_ENM}.

Dual-resolution methods have been successfully applied for the study of several biological systems, including soluble \cite{fiorentini2020ligand} and membrane proteins \cite{alfonso2019multiscale, alfonso2019understanding, fierro2019dual}. However, the available approaches share some common shortcomings: first, the standard modelling of the CG region allows little flexibility in the choice of the CG sites; second, the CG region is usually defined \textit{ad hoc}, and new mappings require a completely new reparameterization of the interactions; third, non-bonded interactions (such as electrostatics) are typically not taken into account in the CG model, thus preventing interactions between different structural domains that might come in close contact during the course of the simulation. CG models with an accurate description of electrostatics have been developed \cite{spiga2013electrostatic, darre2015sirah, souza2021martini}; however, in such cases, the protein is uniformly coarse-grained at a resolution intermediate between the atomistic and one-bead-per-amino acid one, thus limiting the level of coarse-graining and preventing a straightforward coupling between regions at different resolution. These limitations hinder the applicability of standard multiple-resolution models, with detrimental consequences for the {\it in silico} investigation of proteins and their interactions.

In this work we propose a novel approach, dubbed coarse-grained anisotropic network model for variable resolution simulations, or CANVAS, which enables a fast parametrisation of multiple-resolution models. The CANVAS strategy leverages the blurred and approximate nature of coarse-grained models to identify effective sites based on a user-provided input, and determines the interactions among them based on the molecule's structure and all-atom force field, making it unnecessary to run reference simulations. This strategy makes the parametrisation of the model practically instantaneous, and allows the modulation of the system's resolution in a quasi-continuous manner across the structure, from all-atom to (very) coarse-grained. Most notably, the interaction between regions of the system at different resolution (including the solvent) is accounted for and straightforward to set up, allowing the seamless implementation in standard MD software packages (e.g. GROMACS or LAMMPS).

The paper is structured as follows: first, we describe in detail the CANVAS model, focusing on the construction of the multiple resolution representation and on the parameterization of the interactions. A Methods section follows, providing the simulation details. The results of the validation of the CANVAS approach are then presented, by comparing results from all-atom and multiscale simulations of two biomolecules, namely the enzyme adenylate kinase and the IgG4 antibody pembrolizumab, each modelled with three resolution levels. Finally, conclusions and perspectives are discussed.

\section{The CANVAS model} \label{sec: CANVAS-model} 

\begin{figure}[htp]%[htbp]
\centering
\includegraphics[width=\columnwidth]{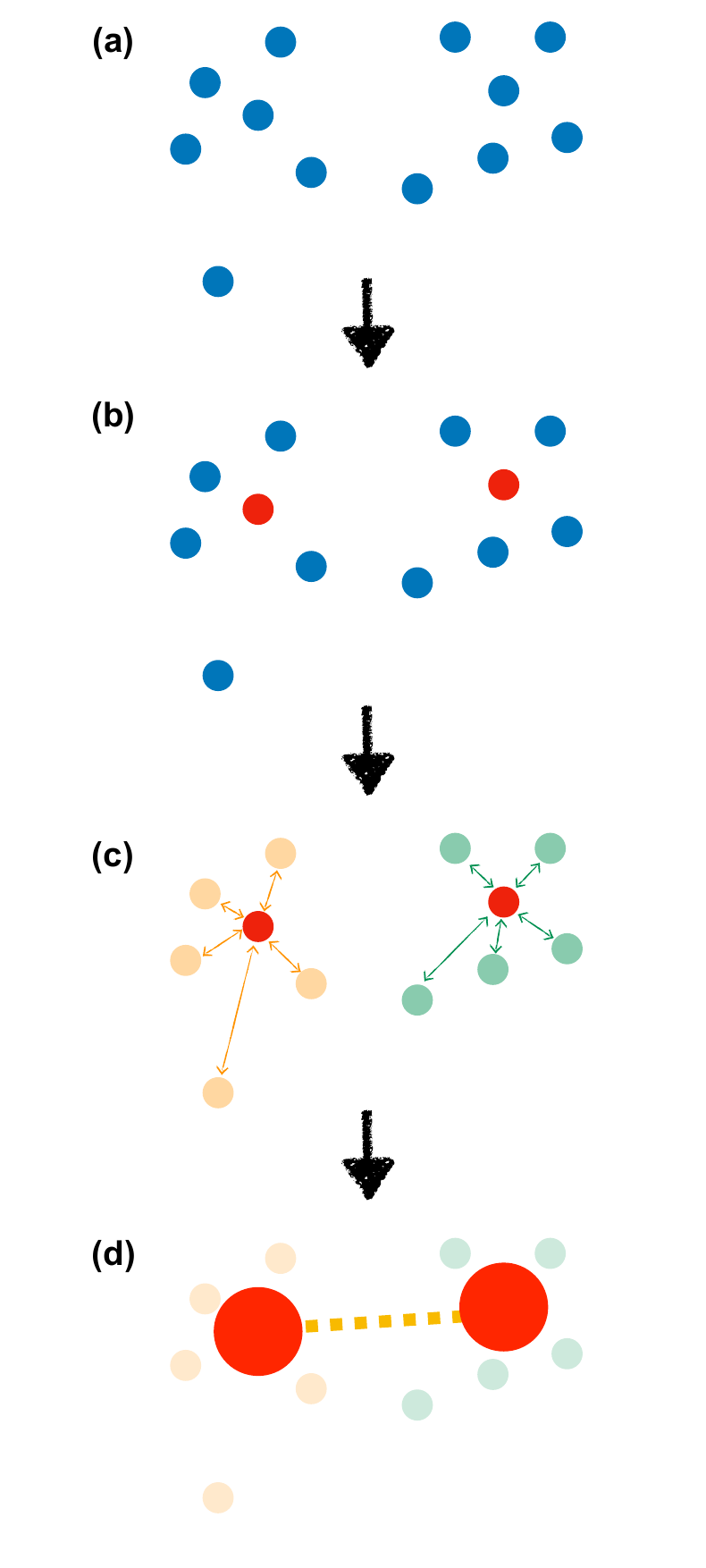}
\caption{Scheme of the decimation process in the low-resolution part of the biomolecule. (a): blue circles show all the atoms in the low-resolution part. (b): choice of the atoms that survive depicted in red. (c): the decimated atoms (light orange and light green) are mapped onto their closest survived atom in terms of Euclidean distance (orange and green arrows), according to the Voronoi tessellation. (d): Each survived atom, shown with a large red circle, is representative of the closest not-survived atoms mapped by it. A harmonic spring, depicted with a dashed yellow line, connects the neighbouring survived atoms.}
\label{fig: Prot-Voronoi}
\end{figure}

In the CANVAS approach to multi-resolution protein modelling, a decimation mapping is implemented for the choice of the interactions sites \cite{giulini2021system}: those atoms included in a user-defined list are retained, while the other ones are discarded. If all atoms of a given subregion of the molecule are retained, the high-resolution atomistic description is employed; on the contrary, regions where atoms are removed are described at a varying level of detail. In lower-resolution regions, the physical properties of the survived atoms are modified so as to incorporate in effective interactions those atoms that have been integrated out (Figure \ref{fig: Prot-Voronoi}). Specifically, each discarded atom is associated to the closest surviving one, and the properties of the latter are determined from those of the group of discarded atoms it represents.

The CANVAS model enables in principle a quasi-continuous modulation of the resolution of a protein or part of it, in that the detail of representation can be gradually reduced from the all-atom level to a very coarse one, possibly lower then a few ($1$ to $3$) amino acids per bead; between highest and lowest resolutions, an arbitrary number of intermediate levels are feasible. In the current implementation, we performed the choice of employing three levels of resolution:

\begin{itemize}
    
    \item \textbf{all-atom (AT)}: the highest level of detail, where all the atoms of a given amino acid are retained; 
    
    \item \textbf{medium-grained (MG)}: intermediate level of detail, where only the backbone atoms of an amino acid are retained, i.e. the carbon alpha CA$_{mg}$, the nitrogen $N_{mg}$ of the amino group, the oxygen $O_{mg}$ and the carbon $C_{mg}$ of the carboxyl group.
    
    \item \textbf{coarse-grained (CG)}: the lowest level of resolution. In the applications presented here, only the C$_\alpha$ atoms of each CG residue are kept, dubbed CA$_{cg}$. 
    
\end{itemize}

The sets of protein residues modelled with an AT, MG, or CG detail are specified by the user and do not change during the simulation, that is, the biomolecule has a time-independent triple resolution. Table \ref{tab:regions} summarizes the survived atoms in each region and their label.

\begin{table}[ht]
\centering
\begin{tabular}{|c|c|c|c|}
\hline
\bfseries{Label} & \bfseries{Region} & \bfseries{Survived atoms per aa} \\
\hline 
&&\\[-0.3em]
at      & fully-AT          & all (CA$_{at}$, N$_{at}$, etc.) \\ \cline{1-3}
&&\\[-0.3em]
mg      & medium-grained    & backbone (N$_{mg}$, CA$_{mg}$, C$_{mg}$,  O$_{mg}$) \\ \cline{1-3}
&&\\[-0.3em]
cg      & coarse-grained    & C$_\alpha$ (CA$_{cg}$) \\ \hline
\end{tabular}
\caption{Description of survived atoms for amino acids (aa) for each level of resolution.}\label{tab:regions}
\end{table}

The first step of the model construction is to identify the region of the system where the chemical details play a crucial role, such that no simplification of the atomistic description is desirable. Residues described at MG and CG resolutions can be either specified from the user or directly identified on the basis of the atomistic residues; in the latter case, the MG region is built by including those residues at a distance of $1$ nm from the AT region, while the rest of the biomolecule is automatically assigned a CG representation.

The AT part is modelled through a standard atomistic force field (in the implementation discussed here, these are Amber99SB-ildn \cite{amber99-ilbn} or CHARMM36m \cite{charmm36m}), where the classical functional form and parameterization of the bonded and non-bonded interactions between atoms are employed. In the MG and CG domains, the potential energy is given by:
\begin{equation} \label{eq:ff_CG}
 \begin{split}
    E & = E_{AA} + E_{harmonic} \\
    & + E_{VdW} + E_{coulomb}.
 \end{split}
\end{equation}

The first term, $E_{AA}$, corresponds to bonded interactions from the atomistic force field, namely chemical bonds, angles, and proper/improper dihedrals:
\begin{equation} \label{eq:amber}
 \begin{split}
    E_{AA} & = E_{bonds} \cdot h \left(r\right) + E_{angles} \cdot h(\theta) \hspace{2mm} + \\
    & + E_{dihedrals} \cdot h(\phi) + E_{improper} \cdot h(\omega).
 \end{split}
\end{equation}

\begin{figure*}[htp]%[htbp]
\centering
\includegraphics[width=0.78\textwidth]{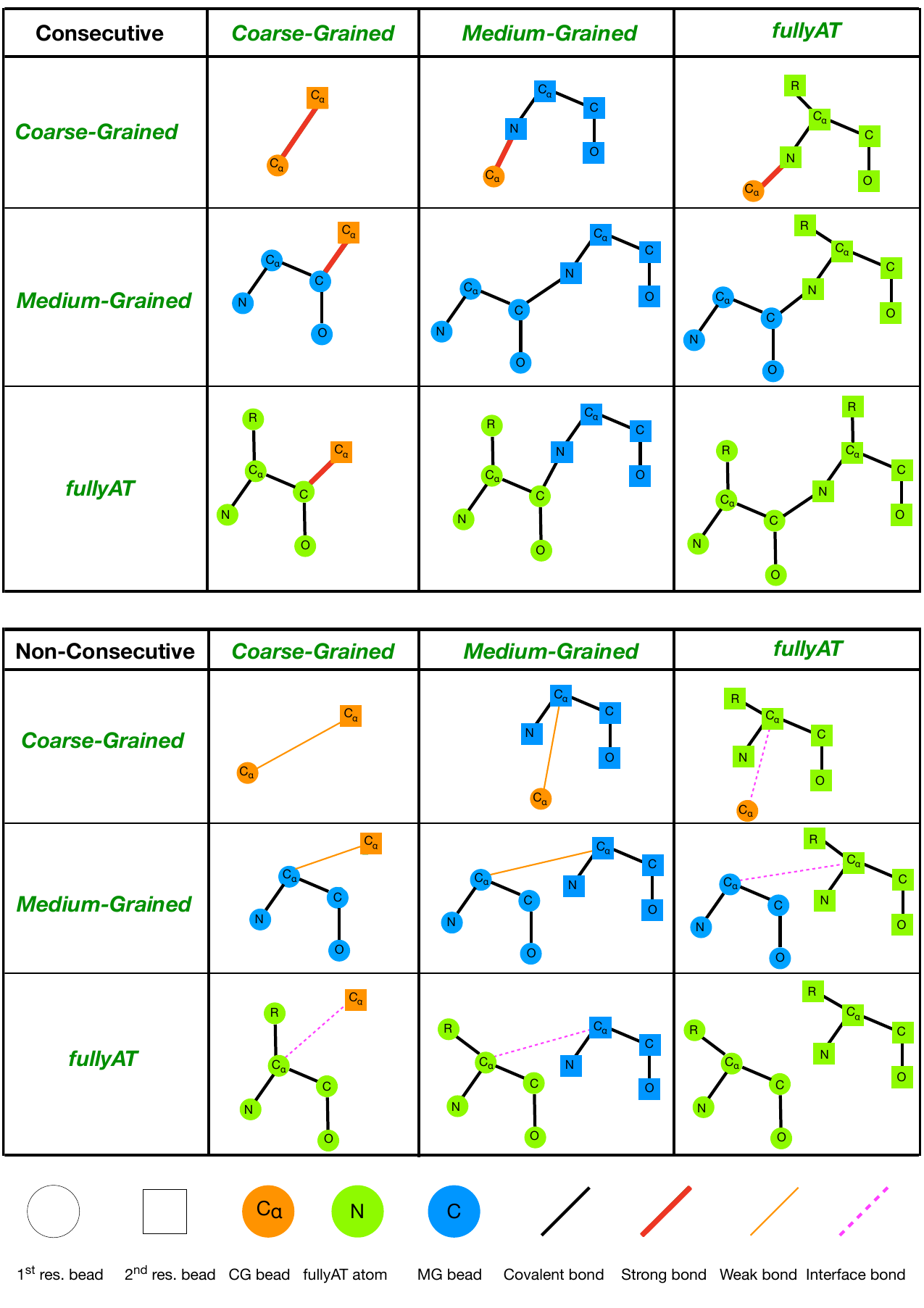}
\caption{Schematic representation of the bonded interactions in the three regions at different resolution. On the top of the figure, only consecutive residues are considered; on bottom, non-consecutive ones. The atoms/beads that belong to the 1st residue are traced with a circle, while those ones that belong to the 2nd residue are sketched with a square. \textbf{R} stands for the side chain. In the figure, hydrogen atoms are ignored for clarity, while being explicitly accounted for in the model. Bonded interactions are represented with different colors and thicknesses according to the spring constant.}
\label{fig: bonds-canvas-pic}
\end{figure*}

Here, $h(r)$, $h(\theta)$, $h(\phi)$, $h(\omega)$ are Heaviside functions taking value $1$ if a bond, angle, dihedral or improper dihedral exists in the atomistic force field for a couple, triplet, or quadruplet of survived atoms. Therefore, stretching, bending, and torsion potentials with their original equilibrium values are possible only if, respectively, the pair, triplet, and quadruplet of atoms (where at least a CG bead is involved) from the all-atom representation of reference are maintained in the MG and CG regions. The second term in Eq. \ref{eq:ff_CG}, $E_{harmonic}$, describes the bonded interactions between and within the low-resolution domains. The bonded connectivity and its parameterization are strictly dependent on the resolution levels employed and on the chemical nature of the retained sites, namely on their \textit{atom type}. In the current implementation, beads are connected by harmonic springs as schematically depicted in Figure \ref{fig: Prot-Voronoi}d and described in detail in Figure \ref{fig: bonds-canvas-pic}. Specifically, the reference bond length corresponds to the distance between the two atoms/beads in the starting structure, while the value of the elastic constant depends on the nature of the bonded particles and their position along the sequence:

\begin{enumerate}
    \item a stiff spring $(k_b)$ is employed for consecutive beads (red line of Figure \ref{fig: bonds-canvas-pic}); its value is $5 \cdot 10^4$ kJ $\cdot$ mol$^{-1} \cdot$ nm$^{-2}$.
    \item A weaker spring $k_{nb}$ is used for non-consecutive C$_\alpha$ beads (CA$_{cg}$ -- CA$_{mg}$, CA$_{cg}$ -- CA$_{cg}$, CA$_{mg}$ -- CA$_{mg}$) whose distance in the reference (native) conformation lies below a fixed cutoff equal to $1.4$ nm (orange line of Figure \ref{fig: bonds-canvas-pic}). Critically, the magnitude of $k_{nb}$ depends on the distance $d$ between the two C$_\alpha$ beads, farther CG units interacting through looser springs. The profile of $k_{nb}(d)$ was obtained through a statistical analysis performed over an ensemble of effective pair potentials acting among non-consecutive C$_\alpha$ atoms in the pembrolizumab antibody; such potentials were extracted {\it via} direct Boltzmann inversion. See Section S1 in the Supporting Information for further technical details.
    \item A second weaker spring $k_{if}$ is employed between an atomistic C$_\alpha$ and a CA bead (CA$_{at}$ -- CA$_{mg}$ or CA$_{at}$ -- CA$_{cg}$) if they do not belong to consecutive residues, and their distance in the reference conformation is less than a fixed cutoff equal to $1.4$ nm (magenta dash line of Figure \ref{fig: bonds-canvas-pic}). The recommended value of $k_{if}$ is $50$ kJ $\cdot$ mol$^{-1} \cdot$ nm$^{-2}$ in order to guarantee the appropriate degree of flexibility.
    
\end{enumerate}

We stress that, if the two survived atoms taken into account are connected by a covalent bond in the fully-atomistic representation, the latter replaces $E_{harmonic}$ (black line of Figure \ref{fig: bonds-canvas-pic}). Similarly, bending and torsion potentials with their original atomistic parameterization are maintained if the triplet and quadruplet of atoms (where at least a CG bead is involved) in the all-atom representation of reference are retained in the coarse regions. Rescaled non-bonded 1-4 interactions are introduced only in the AT region.
In addition, in order to guarantee the correct degree of flexibility in multidomain proteins, no bond is introduced between those beads that are close in space in the starting configuration but belong to distinct structural domains; the latter can be defined either on the basis of the knowledge of the system, or through appropriate algorithms developed to decompose protein structures in rigid subunits~\cite{potestio2009coarse, bernhard2010optimal}. The indices of the residues belonging to each domain are specified by the user in an optional input file.

\begin{figure}[htp]%[htbp]
\centering
\includegraphics[width=\columnwidth]{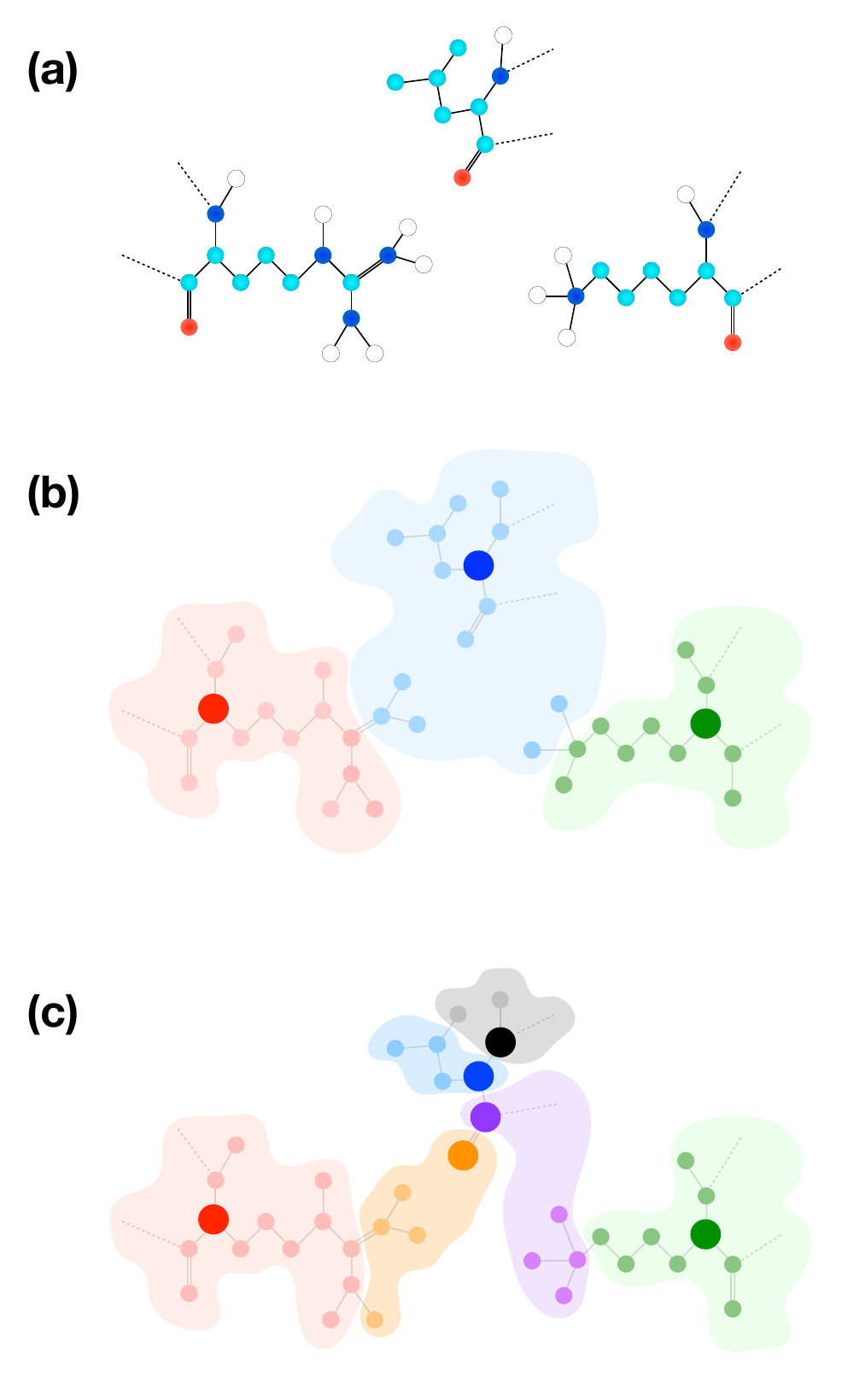}
\caption{(a): schematic representation of three amino acids arginine, leucine and lysine (from left to right). Dashed black lines represent the peptide bonds between two residues. The aliphatic hydrogens are not displayed for simplicity. (b): all the three amino acids are modelled as CG, where only the C$_\alpha$ atom CA$_{cg}$ (red, blue, and green bigger circles) are retained. The other atoms are decimated and mapped onto the closest survived atom (shown in pink, light blue, and light green). A bead is not necessarily representative of atoms belonging to the same residue, since the grouping criterion is merely based on euclidean distance. (c): arginine and lysine are modelled in CG (red and green bigger circles), whereas the leucine is described in MG (CA$_{mg}$ in blue, N$_{mg}$ in black, C$_{mg}$ in violet, and O$_{mg}$ in orange).}
\label{fig: new-leu-arg-lys}
\end{figure}

Finally, E$_{VdW}$ and E$_{coulomb}$ in Eq. \ref{eq:ff_CG} are the Van der Waals and Coulomb non-bonded contributions to the potential energy between nodes. For the AT region, standard force-field parameters are taken, while in the MG and CG regions, the charge and Lennard-Jones parameters of each bead are computed from the average properties of the neighbouring atoms, as defined through a procedure akin to a Voronoi tessellation \cite{voronoi1, voronoi2, voronoi3}. First, a Voronoi cell is defined by associating the decimated atoms (blue circles of Figure \ref{fig: Prot-Voronoi}b) to the closest survived atom (in terms of Euclidean distance $\ell$), which is now treated as a CG bead (Figure \ref{fig: Prot-Voronoi}c, and Figure \ref{fig: new-leu-arg-lys}b-c). We underline that, since a geometric criterion is employed to group atoms, the resulting bead is representative of atoms that could also belong to separate residues, as schematically shown in Figure \ref{fig: new-leu-arg-lys}. For this reason, the protein's starting structure plays a relevant role in the Voronoi tessellation, since the relative orientation of side-chains might influence the construction of the cells. Therefore, it is important that the structure employed as a reference for coarse-graining is minimized and equilibrated. For the same reason, the Voronoi tessellation-based coarse-graining procedure is strongly dependent on the starting structure, and we can expect the relative arrangements of secondary elements to be preserved during the simulation. If conformational changes are desirable, a careful distribution of the different degrees of resolution along the structure is required, and a more informed partition of the system should be done with explicit input from the user.

After the definition of the Voronoi cells, non-bonded potential parameters are computed for each CG bead. Specifically, for a mapping that retains $N$ atoms out of $n$:
\begin{itemize}
    \item the charge $Q_{_I}$ is defined as the algebraic sum of the charges $q_i$ of the atoms it represents:
    \begin{equation}\label{eq: Q}
        Q_{_I} \equiv \sum_{i \in I} q_i.
    \end{equation}

    \item The diameter $\sigma_{_I}$ is twice the gyration radius $R_g$:
    \begin{eqnarray}\label{eq: sigma}
        &&\sigma_{_I} \equiv 2 \cdot R_g\\ \nonumber
        &&\mbox{where:}\\ \nonumber
        &&R_g^2 = \frac{1}{N}\cdot\sum_{i=1}^N\left|\textbf{r}_i - \textbf{r}_{\text{cog}}\right|^2\\ \nonumber
        &&\textbf{r}_{\text{cog}} = \frac{1}{N}\sum_{i=1}^N \textbf{r}_i.
    \end{eqnarray}
    Here, $\textbf{r}_i$ are the coordinates of each atom, whereas $\textbf{r}_{\text{cog}}$ corresponds to the coordinates of the center of geometry of the group.

    \item $\epsilon_{_I}$ is the geometric average of the $\epsilon_i$ values of the atoms it represents:
    \begin{equation}\label{eq: epsilon}
        \epsilon_{_I} \equiv \prod_{i \in I} \epsilon_i^{\frac{1}{N}}.
    \end{equation}

\end{itemize}

As opposed to the network of bonded interactions, where a predefined set of parameters is employed, the non-bonded part is automatically constructed on the basis of the properties of the retained sites, independently of the level of resolution and the bonded connectivity between them. The combination rule used to determine, from these parameters, Lennard-Jones interactions between CG beads is the same as the one employed by the atomistic force field in the high-resolution region; namely, it is based on the Lorentz-Berthelot rules for both the Amber and CHARMM force fields. In addition, in the case of interactions between non-consecutive coarse-grained sites, non-bonded interactions are fully accounted for, while non-bonded interactions are switched off in the case of bonds involving atoms in the high-resolution region, as in the standard atomistic description.

We stress that, since CG beads in the CANVAS representation may not be representative of a single residue, a direct residue-based analysis can not be performed. This is a specific feature of the CANVAS approach: the latter, in fact, was conceived to be easily generalized to very coarse mappings, where one bead is representative of more than one residue; or for inhomogeneous mappings, where the retained low-resolution sites are distributed throughout the protein independently from the residue at which they originally belong (so that some residues might be represented by one or more beads, while others might be discarded completely). In such a case one can use mappings that are different from the intuitive, chemistry-based ones, but that are the most efficient in preserving the information contained in the all-atom protein representation \cite{diggins2018optimal, giulini2021system}.

The code and examples of input files for simulating a system with the CANVAS model are freely available at \href{https://github.com/potestiolab/canvas}{https://github.com/potestiolab/canvas}. The code consists of two python scripts: the first one (\textit{block.py}) has the purpose of creating the list of survived atoms with their relative labels (AT, MG, or CG); the second script (\textit{CANVAS.py}) returns the input files needed for simulating a solvated biomolecule in LAMMPS or GROMACS, according to the choice made by the user. The mandatory arguments for the successful execution of the code are the list of survived atoms, the coordinate file (.gro) and the topology file (.top) of fully-atomistic representation. A detailed description of the other parameters (mandatory and optional) and a tutorial for the construction and simulation of a CANVAS model, starting from the atomistic representation, are available on the same repository.

\section{Materials and methods}\label{sec: simul-details}

The two systems employed in the present work as a test bed for the CANVAS model are the enzyme adenylate kinase \cite{ake1, ake-struct, 4ake2} and the antibody pembrolizumab \cite{scapin2015structure}.

\begin{figure}[htp]%[htbp]
\centering
\includegraphics[width=\columnwidth]{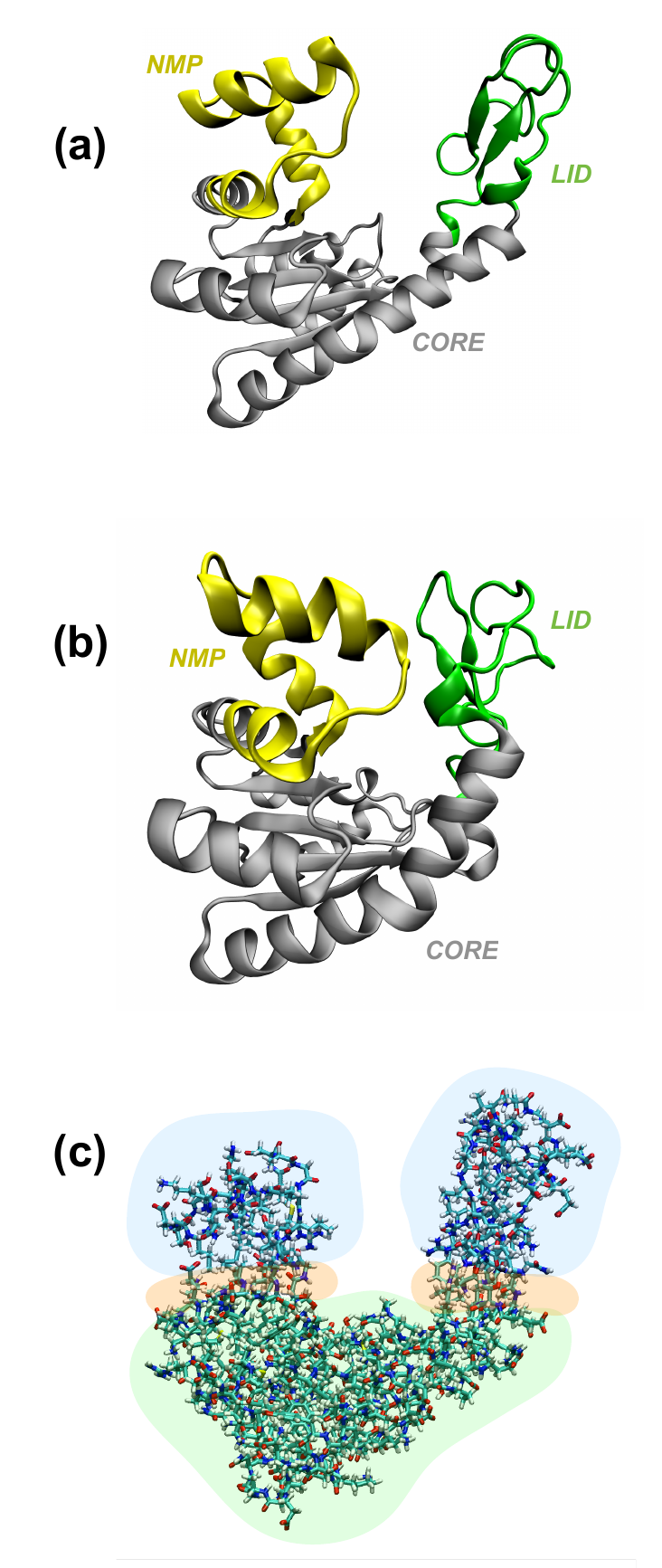}
\caption{Fully atomistic representation of ADK. In particular, (a) and (b) show the open and the compact conformation of the protein, respectively, in terms of secondary structure. LID, NMP, and CORE domains are depicted in green, yellow, and grey. (c) displays a schematic representation of the reference structure of ADK before conversion from the all-atom representation to the CANVAS one. Specifically, the CORE of protein, modelled atomistically, is depicted in green; the part that is described in MG is shown in orange; the remainder, which is going to be coarse-grained, is shown in blue.}
\label{fig: ake-divisions}
\end{figure}

Adenylate kinase (ADK) plays a critical role in maintaining the energetic balance in the cell, interconverting adenosine diphosphate (ADP) molecules into adenosine monophosphate (AMP) and adenosine triphosphate (ATP) \cite{ionescu2019adenylate}. The structure of ADK can be partitioned in three domains, called core, lid, and NMP, and two distinct nucleotide binding sites, as shown in Figure\ref{fig: ake-divisions}a-b.

\begin{figure}[htp]%[htbp]
\centering
\includegraphics[width=\columnwidth]{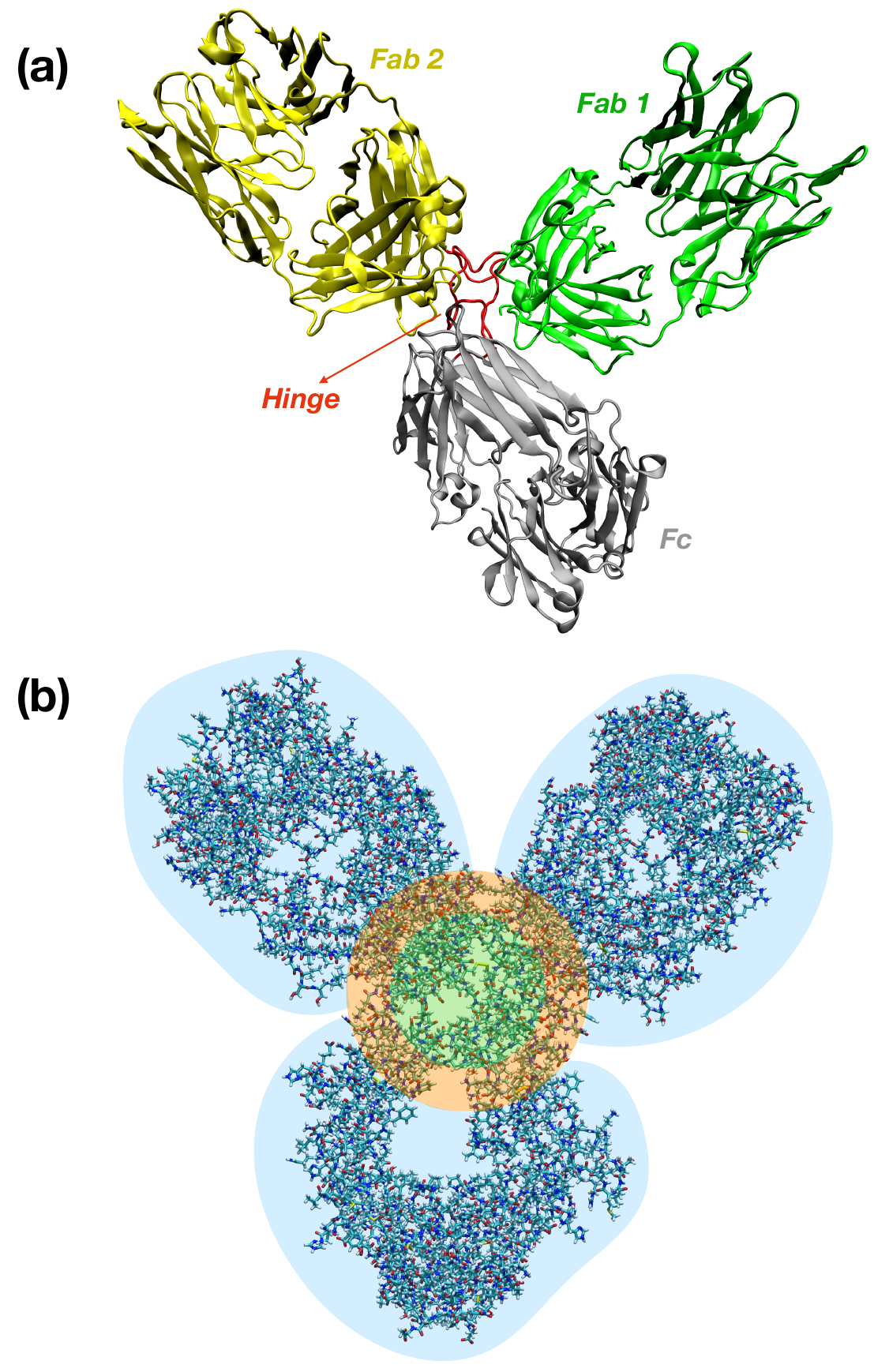}
\caption{(a): Graphical representation of the crystallographic conformation of pembrolizumab in terms of secondary structure. Fab 1, Fab 2, Fc, and the hinge are depicted in green, yellow, grey, and red, respectively. (b): Schematic representation of the 4$_A$ reference conformation of pembrolizumab before conversion from the all-atom to the multi-resolution. In particular, in green it is depicted the hinge of biomolecule, modelled atomistically; in orange it is shown the part that is going to be described as MG; the remainder, shown in blue, is going to be represented as CG.}
\label{fig: antibody-plus-division}
\end{figure}

The second system used here as a test case, pembrolizumab, is a humanized IgG4 antibody consisting of four chains, covalently bound by disulfide bonds (Figure \ref{fig: antibody-plus-division}a).
Pembrolizumab -- which is the generic name for the trade drug name Keytruda\textsuperscript{\textregistered} -- is currently used in immunotherapy as an anticancer drug \cite{ivashko2016pembrolizumab}. Its antigen is the programmed cell death protein 1 (PD-1), expressed on the membrane of T cells, B cells, and natural killer cells; the formation of the high-affinity complex between the antibody and its antigen prevents the binding of PD-1 with the programmed cell death receptor ligands PD-L1 and PD-L2, which would lead to a suppression of the anti-tumor activity of T cells \cite{chen2020looking}.

The reference structures employed for the construction of the multiscale models were obtained from equilibrated all-atom simulations. Specifically, the crystallographic ADK structure (PDB ID: 4AKE) was solvated in an aqueous solution at $0.15$M NaCl concentration; following energy minimization, the system was equilibrated for 125~ns in the NPT ensemble, using the Parrinello-Rahman barostat \cite{Parrinello} with a time constant of $2.0$ ps at $1$ bar and the Langevin thermostat \cite{langevin-thermostat} to keep the temperature at 300 K. The all-atom simulation was extended for 500 ns, on which the analyses were performed. For the CANVAS simulation, the equilibrated structure was placed in a cubic simulation box of $9.1$ nm per side and solvated in an aqueous solution at $0.15$M NaCl concentration.

The reference structures of pembrolizumab are given by the representative conformations sampled from four all-atom $500$~ns-long simulations of the antibody in the apo form, after clustering the frames on the basis of their structural similarity. Each of these atomistic simulations was started from the PDB crystallographic structure of the deglycosilated antibody (PDB ID: 5DK3), after modelling of the missing residues; for more details on the all-atom simulation protocol, the reader is referred to Ref. \cite{tarenzi2021communication}. A CANVAS simulation is started from each representative conformation of the antibody, for a total of 6 different runs; this choice is dictated by the large conformational variability of the molecule, and the peculiar properties of each conformational basin. The CANVAS models of the representative structures are placed in a cubic simulation box of $17.9$ nm per side and are solvated in a $0.15$M NaCl aqueous solution.

For both ADK and pembrolizumab, the force field employed was Amber99SB-ildn \cite{amber99-ilbn} and the water model was TIP3P \cite{Klein_JChemPhys_1983-tip3p}. Furthermore, for the sake of assessing the validity of the approach independently of the specific all-atom force field employed, 10 ns long CANVAS simulations of ADK have been performed with Charmm36m force field, using as MD software programs GROMACS and LAMMPS; the results of these tests are provided as supporting information in Figure S2. CANVAS systems were prepared starting from the representative structures obtained from the atomistic simulations, after energy minimization with the steepest descent algorithm and $100$ ps of NVT equilibration. The temperature is kept constant at 300 K by means of the Langevin thermostat \cite{langevin-thermostat}. In the NPT production run, the Parrinello-Rahman barostat is employed, as described above. The integration step is $2$ fs. The calculation of electrostatic interactions is performed in all cases by using the reaction-field method \cite{reac-field1, reac-field2} with a dielectric constant of $\epsilon = 80$ and a cutoff of $2.5\cdot\sigma_{\text{max}}$; here, $\sigma_\text{max}$ is the maximum value of $\sigma$ among all the beads of the system.
In order to validate the choice of the AMBER force field in combination with the reaction-field method making use of the previous set of parameters, we also performed an atomistic MD simulation using PME for the description of electrostatic interactions, with a dielectric constant $\epsilon = 1$ and a cutoff of 1.0 nm. The comparison is performed in terms of RMSF between the two all-atom trajectories of ADK, as shown in Figure S3. We observed that the trends of fluctuations are consistent with each other, providing comforting evidence that the AMBER model can be safely employed with reaction field.
The SETTLE \cite{Kollman_JCompChem_1992-SETTLE} and RATTLE \cite{Andersen_JComputPhys_1983-RATTLE} algorithms for rigid water and rigid bonds containing hydrogen have been used. The length of the CANVAS simulations is 500 ns for ADK and 200 ns for each antibody system. All simulations are carried out with GROMACS 2018 \cite{manual_grom_2018}. We stress here that the usage of an explicit solvent, while guaranteeing the highest level of accuracy of the model in the atomistic region, makes the computational cost of the simulation essentially identical to that of a fully atomistic model.In Table \ref{tab: performance-ADK} we provide a quantitative comparison of the performance of 500 ns long ADK simulations run on a 48-cores single node. These show how the CANVAS simulation is slightly faster (about 1.05 times) than the atomistic one when using the reaction-field electrostatic method and same cutoff. Moreover, as expected, the all-atom simulation employing the reaction field is faster -- about twice  -- than the corresponding one when using PME. One of the long-term targets in the development of variable-resolution models is the boost of computational efficiency through the reduction of the number of model particles; here, however, we apply the multiscale representation for the biomolecule alone, since the combined usage of multiple-resolution models of the protein {\it and} of the solvent would lead to ambiguities in the validation and in interpretation of the outcomes. The usage of CANVAS in combination with computationally efficient models of the solvent (e.g. implicit solvent \cite{onufriev2019generalized, chen2021machine} or adaptive resolution simulation schemes \cite{praprotnik2007adaptive, Donadio_PRL_2013-hadres_molliq, tarenzi2017open}) will be the object of future work.

\begin{table}
\centering
\begin{tabular}{|c | c | c | c|} 
 \hline
 \textbf{Method} & \textbf{Resolution} & \textbf{Cutoff [nm]} & \textbf{Performance} \\
 \hline\hline
 Reaction Field & all-atom  &  1.000  & 87.90 ns/day \\
 Reaction Field & all-atom  &  1.698  & 32.14 ns/day \\ 
 Reaction Field & CANVAS    &  1.698  & 33.75 ns/day \\
 PME            & all-atom  &  1.000  & 43.04 ns/day \\
 PME            & all-atom  &  1.698  & 19.17 ns/day \\
 \hline
\end{tabular}
\caption{Comparison of the time performance for ADK simulations run on 48 cores, single node, for different electrostatic methods, interaction cutoffs, and resolution. As expected, those employing the PME are about twice as slow as those employing the reaction-field (RF) method for all-atom simulation and cutoff of 1.0 nm. The CANVAS simulation is slightly faster than all-atom one, when using the RF method and a cutoff of 1.698 nm. The latter value corresponds to $2.5\cdot\sigma_{\text{max}}$ for the ADK starting configuration when constructing the CANVAS model.}
\label{tab: performance-ADK}
\end{table}

The analysis of fluctuations was performed with VMD molecular visualization program \cite{VMD}. In particular, the root-mean-square deviation (RMSD) was computed through the \textit{RMSD Trajectory Tool} considering the sole C$_\alpha$ atoms. The root mean square fluctuations (RMSF) were computed by means of an in-house tkl script. The radii of gyration were computed with \textit{gmx gyrate}, while the solvent-accessible surface area was computed with \textit{gmx sasa}. The principal component analysis (PCA) and the calculation of the root mean square inner product (RMSIP)~\cite{amadei1999convergence} between the essential subspaces from atomistic and CANVAS simulations were performed with the Python module MDAnalysis. The calculation of the electrostatic potential was performed with the online adaptive Poisson-Boltzmann solver (APBS) \cite{jurrus2018improvements}, after the creation of an input PQR file that, in the case of the multiscale model, includes the radii and charges as computed with the CANVAS protocol. Protein visualization and rendering was performed with VMD \cite{VMD}, while the plots were created with Xmgrace and Python libraries.

\section{Result and Discussion}\label{sec:results}

In this section we compare results from the atomistic and CANVAS simulations for both ADK and pembrolizumb, in order to assess the validity of the proposed multiscale model. In the case of pembrolizumab, the comparison is performed between the six CANVAS simulations and the corresponding ensembles of structurally homogeneous configurations obtained through a clustering of all-atom simulation frames, see Tarenzi {\it et al.} \cite{tarenzi2021communication}.

\subsection{Adenylate kinase} 

The ADK protein exists in two main conformations, required for the catalytic activity of the enzyme: a fully open one, where the LID and the NMP domains are separated from each other, thus exposing the binding site; and a closed one, which is stabilized by the presence of the substrate and allows for the enzymatic reaction to take place~\cite{onuk2017effects}. In the all-atom simulation, ADK samples both the open conformation, which corresponds to the starting structure (Figure \ref{fig: ake-divisions}a), and a more compact one (Figure \ref{fig: ake-divisions}b), where the distance between the LID and NMP arms is substantially reduced. This partially-closed conformation of ADK in the apo state was already observed experimentally \cite{henzler2007intrinsic} and in previous MD simulation studies \cite{li2015mapping, wang2020exploring}. However, we do not observe a complete transition between the open and the fully closed states, as expected from the absence of the substrate and from the long timescale of the process (of the order of $\mu$s-ms \cite{zheng2018multiple}); indeed, the computed distance between the C$_\alpha$ atoms of residues A55 and V169, previously used to discriminate the two conformational states both in experiments and simulations \cite{whitford2007conformational}, is consistent with the open state for the whole duration of the trajectory (Figure S4 in the Supporting Information).

\begin{figure*}[htp]%[htbp]
\centering
\includegraphics[width=\textwidth]{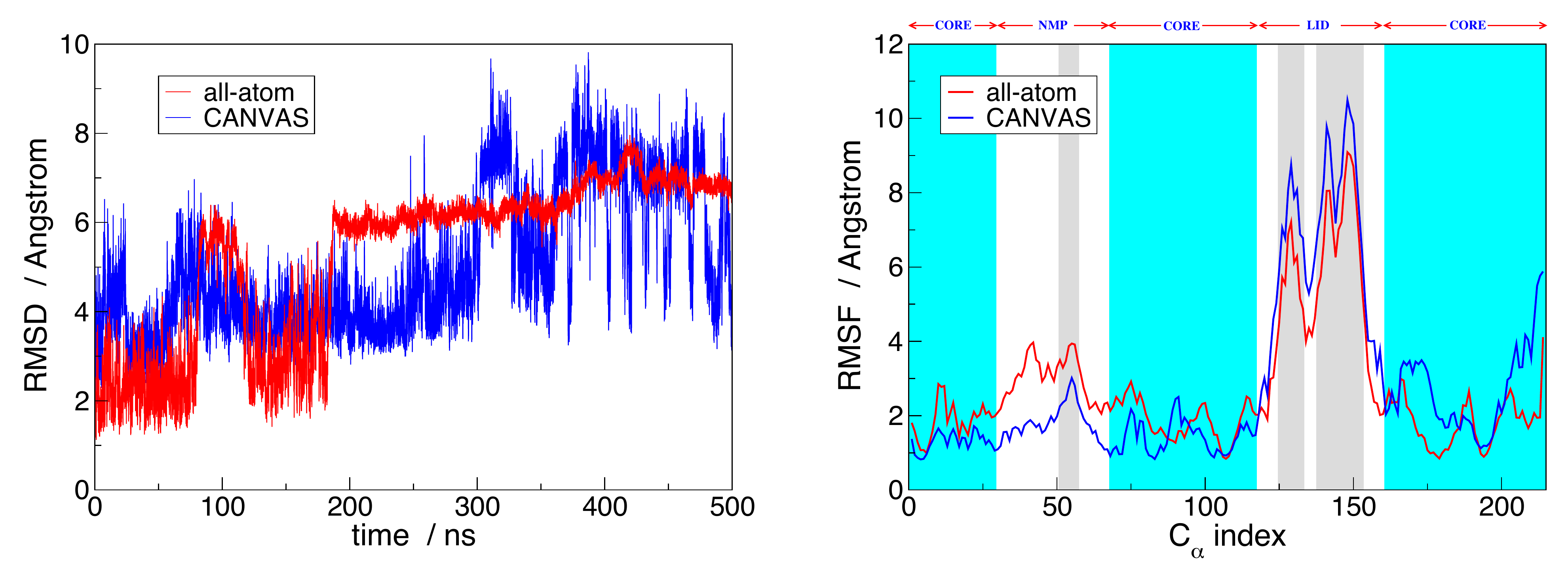}
\caption{(a): RMSD of all ADK C$_\alpha$ comparing the all-atom simulation (red line) and the CANVAS one (blue line). The presence of two different states, one corresponding at about 3 \AA~and the second one close to 6 \AA, are indicative of open and compact conformations, respectively. (b) RMSF for each C$_\alpha$ of ADK from the all-atom simulation (red line) and the CANVAS one (blue line). The cyan area corresponds to the CORE domain, which is described atomistically, while the grey and the white regions correspond to the parts of system (LID and NMP domains) modelled in CG and MG, respectively. Videos of the atomistic and CANVAS trajectories are provided on the Zenodo repository \href{https://doi.org/10.5281/zenodo.7225086}{https://doi.org/10.5281/zenodo.7225086}.}
\label{fig: rmsd-rmsf-4ake-at-canvas}
\end{figure*}

The evolution of the protein between the two aforementioned conformations can be quantified during the simulation in terms of the RMSD of all C$_\alpha$ atoms with respect to the initial frame, which corresponds to the equilibrated structure of ADK in the NPT ensemble (Figure \ref{fig: ake-divisions}a). Since the latter is in the open conformation, higher RMSD values are indicative of closer structures. The resulting plot is shown with a red line in Figure \ref{fig: rmsd-rmsf-4ake-at-canvas}a: as expected, two states are clearly visible: one corresponding to $3$ \AA, and the second one around $6$ \AA. The compact conformation (higher RMSD values) is attained for a few nanoseconds after $80$ ns, it reappears subsequently after $200$ ns, and remains there until the end of the simulation.

Consistently with the previous analysis, the red line of Figure \ref{fig: rmsd-rmsf-4ake-at-canvas}b shows the RMSF for each C$_\alpha$, computed with respect to the average structure: we can notice that the atoms constituting the protein arms, i.e. the LID and NMP domains (indexes 118-160 and 30-67) have wider fluctuations with respect to the CORE.

Since the open/closed transition is determined by the relative orientation of LID and NMP with respect to the CORE, the latter is modelled at high-resolution in the CANVAS simulation, with the aim of retaining a realistic degree of flexibility of the hinge. In contrast, the LID and the NMP domains are described using two levels of resolution, i.e. MG and CG. We recall that all residues whose distance is less than $1$ nm with respect the closest all-atom residues are described as MG, in order to guarantee a \textit{smooth} transition between highest and lowest level of resolution. A schematic representation is shown in Figure \ref{fig: ake-divisions}c.

\begin{figure}[htp]%[htbp]
\centering
\includegraphics[width=\columnwidth]{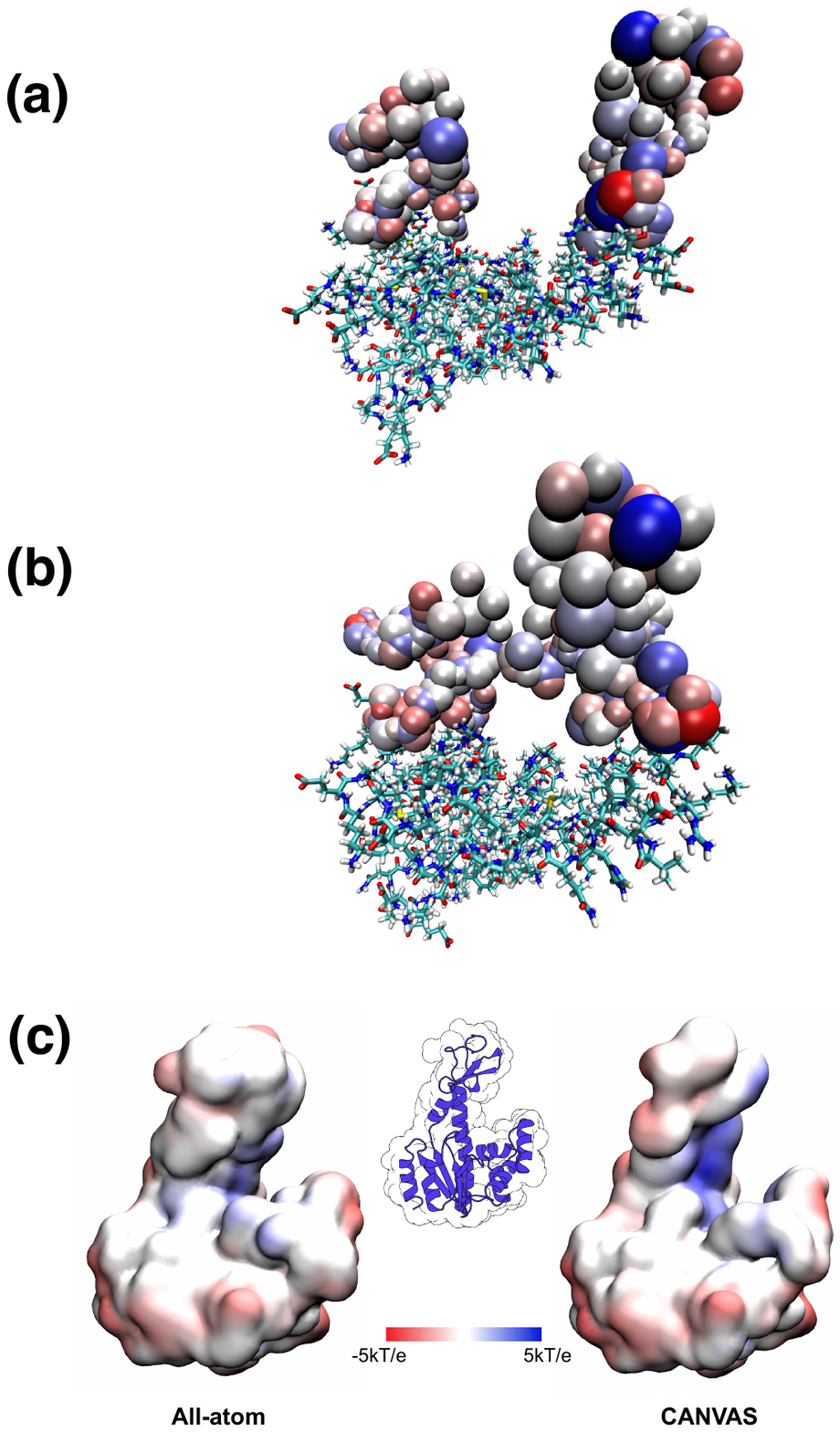}
\caption{CANVAS representations of ADK, where the all-atom region is described in licorice and the MG/CG beads as VdW spheres. The diameter of each bead is given by the value of $\sigma$, while its color is dependent on the value of the charge: white spheres are indicative of neutral charge, while blue and red beads correspond to positive and negative charges, respectively. (a) shows the open conformation of the protein, while (b) the more compact one. (c) displays the electrostatic potential calculated with the adaptive Poisson-Boltzmann solver (APBS) for the all-atom and CANVAS representations of the starting ADK structure, mapped on the protein surface.}
\label{fig: 4ke-canvas-conformation-apbs}
\end{figure}

The CANVAS simulation shows two main protein conformations, in analogy with the all-atom simulation: the open one, as depicted in Figure \ref{fig: 4ke-canvas-conformation-apbs}(a), and the compact one as displayed in Figure \ref{fig: 4ke-canvas-conformation-apbs}(b). The interconversion between the two conformations is monitored, in analogy with the fully-atomistic simulation, by calculating the RMSD of the C$_\alpha$ atoms (CA$_{at}$, CA$_{mg}$, CA$_{cg}$) with respect to the reference frame. The resulting curve is shown in blue in Figure \ref{fig: rmsd-rmsf-4ake-at-canvas}a. The comparison between the all-atom and multi-resolution RMSD shows that the CANVAS model reproduces well the conformational changes observed in the fully atomistic system, allowing the protein to transition between the two basins more frequently than the all-atom reference. In order to assess whether the two sampled states are structurally similar in both simulations, we performed a clustering analysis on the all-atom and CANVAS trajectories, using the RMSD with respect to the starting structure as a distance measure. From the two clusters obtained (corresponding to the fully open and to the compact conformation), the central structures are extracted; representative conformations belonging to the same state are then compared between the atomistic and multiscale cases (Figure S5), and the RMSD between them was calculated. The resulting RMSD values are of 3.7 \AA~for the open conformations, and 5.5 \AA~for the compact ones; a visual inspection of the representative structures reveals that these deviations are mostly limited to the flexible and disordered regions of the protein, while the overall conformational state is the same in the atomistic and multiscale case. Conversely, the comparison of closed and open structures shows larger deviations: the RMSD between the open atomistic and CANVAS compact representative conformations is 7.4 \AA~, while the RMSD value is 5.8 \AA~when comparing the compact atomistic and CANVAS open representative conformations. Next, we looked into the fluctuation of each C$_\alpha$ in the all-atom part, and each CA bead (CA$_{mg}$, CA$_{cg}$) in the MG and CG ones (whose position is the same of the corresponding C$_\alpha$ atoms in all-atom representation), as depicted with blue line in Figure \ref{fig: rmsd-rmsf-4ake-at-canvas}b. Also in this case, for both all-atom and lower-resolution regions the fluctuations of C$_\alpha$ atoms are comparable with those from the atomistic simulation. The comparison of fluctuations has also been performed independently on the sets of frames extracted from the atomistic and CANVAS trajectories after the RMSD-based clustering; the resulting RMSF is plotted in Figure S6 of the Supporting Information, and it shows a similar trend in the two cases.

As explained in the description of the model, the values of $Q$, $\sigma$ and $\epsilon$ for each low-resolution bead are different depending on the number and type of the atoms that are mapped onto it. Figure \ref{fig: 4ke-canvas-conformation-apbs} shows the two conformations where each CG bead is colored according to its charge, and whose size is based on the $\sigma$ values. The partial charges assigned to each MG and CG bead, in addition to those assigned to each atom by the atomistic force-field, were used to compute the electrostatic potential with the adaptive Poisson-Boltzmann solver (APBS) \cite{jurrus2018improvements}. The protein surface, colored according to the mapped potential, is represented in Figure \ref{fig: 4ke-canvas-conformation-apbs}.c for both the fully atomistic and the CANVAS case; the comparison shows that the electrostatic patches are conserved in the multi-resolution representation.

\begin{figure*}[htp]%[htbp]
\centering
\includegraphics[width=0.8\textwidth]{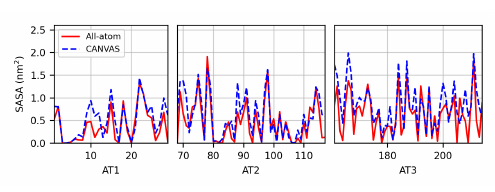}
\caption{Per-residue values of SASA, computed for the atomistic region of the ADK. The AT region is composed of three segments of consecutive residues, denoted AT1, AT2 and AT3.}
\label{fig:sasa-ake}
\end{figure*}

To check the accuracy in the description of the AT region in the CANVAS model, we computed the average solvent-accessible surface area (SASA) for each atomistic residue, comparing the results with the values obtained from the atomistic simulation (Figure \ref{fig:sasa-ake}). The results are in good agreement; the slightly larger SASA values for some residues in the CANVAS simulation might be ascribed to the fact that in the fully atomistic case the protein spends a larger portion of the trajectory in the compact state, where the solvent accessibility of a number of residues is reduced.

\subsection{Pembrolizumab}

In Tarenzi {\it et al.} \cite{tarenzi2021communication}, four all-atom simulations have been performed starting from the PDB crystallographic structure of the full-length antibody, for a total simulation time of 2$\mu$s. The antibody conformations sampled from the MD simulations are grouped into six cluster, on the basis of their structural similarity. The representative structures of the different clusters are shown in Figure S7, and labelled $0_A$, $1_A$, $2_A$, $3_A$, $4_A$, $5_A$ according to the increasing value of the protein's average radius of gyration. The conformations differ mainly in the relative orientation of the Fab and Fc domains, which can get in close contact thanks to the flexibility of the hinge region; the latter includes two 18-residue long disordered segments, bridged by two disulfide bonds.

The six representative pembrolizumab structures are taken as starting conformations for six CANVAS simulations. Since the variation in the relative arrangement of Fab1, Fab2 and Fc domains is made possible by the disordered hinge region, the latter is described atomistically, while the three large domains are modelled with lower levels of detail. In particular, those residues whose distance is less than $1$ nm with respect to the closest fully-atomistic ones are described as MG, while the rest is represented as CG. A schematic representation is given in Figure \ref{fig: antibody-plus-division}b. 

Deviations from the starting structure along the simulations are plotted in Figure S8 in terms of RMSD, and compared with the average RMSD of the atomistic frames falling within the same conformational cluster. Both atomistic and CANVAS deviations were computed with respect to the same reference structure. For the majority of the conformational clusters, the RMSD values from the CANVAS simulation fall within the error bar of the atomistic reference. However, Figure S8 suggests also that the CANVAS representation of pembrolizumab is slightly more rigid than the fully atomistic case; conversely, the atomistic conformations falling within the same cluster appear more heterogeneous, hence their largest values of RMSD with respect to the representative structures.

\begin{figure*}[htp]%[htbp]
\centering
\includegraphics[width=\textwidth]{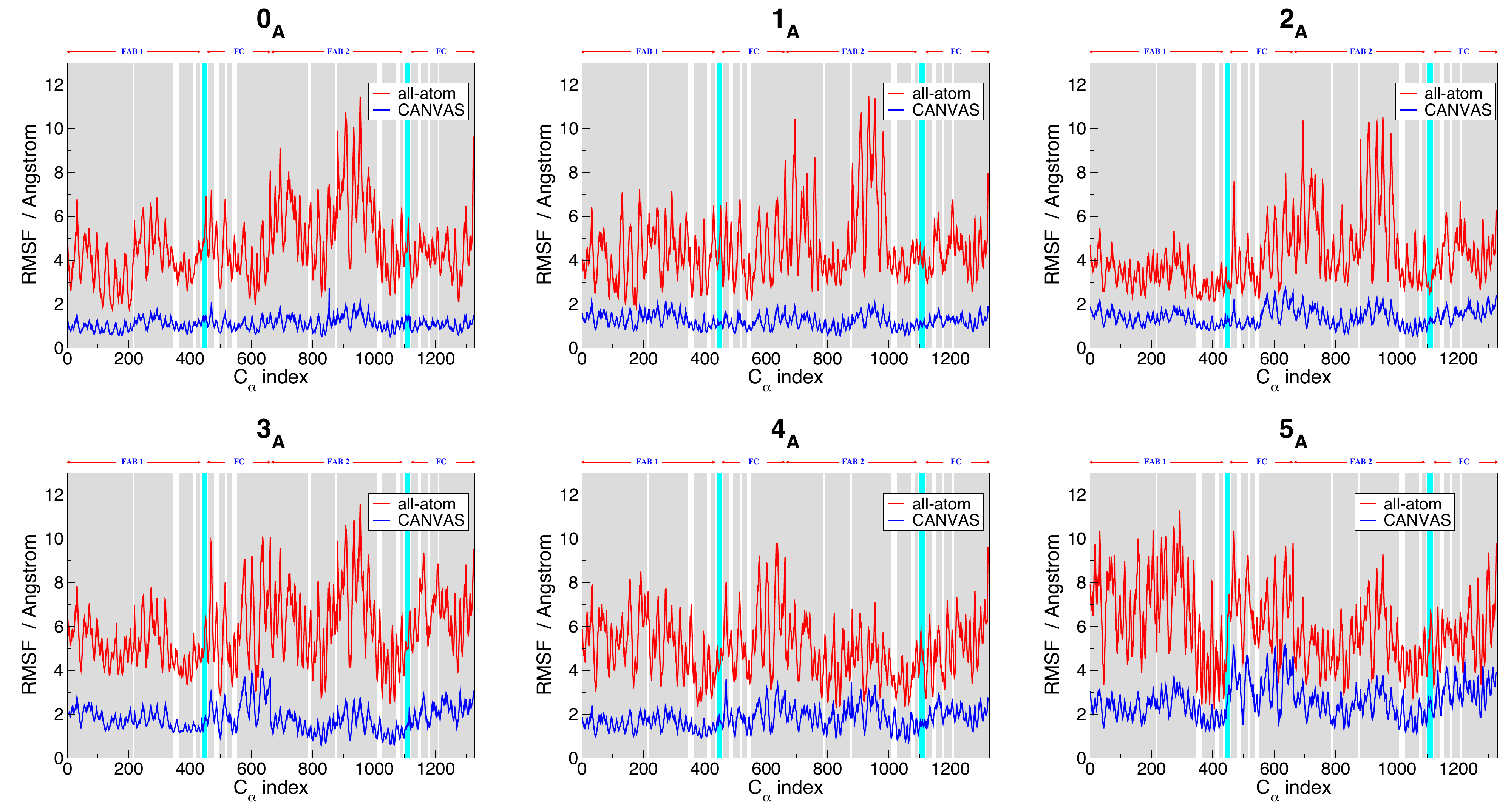}
\caption{Root Mean Square Fluctuation (RMSF) computed on C$_\alpha$ atoms of each apo form of pembrolizumab, for all-atom (red lines) and CANVAS simulations (blue lines). The cyan slabs correspond to the hinge region described atomistically in the CANVAS model, while the grey and the white regions correspond to the parts of system modelled as CG and MG respectively. The $x$-axis corresponds to the C$_\alpha$ indexes.}
\label{fig: rmsf-all-canvas-at-apo}
\end{figure*}

The average residue fluctuations were evaluated by computing the RMSF of each C$_\alpha$ in the all-atom part, and each CA bead (CA$_{mg}$, CA$_{cg}$) in the MG and CG ones (whose position is the same of the corresponding C$_\alpha$ atoms in all-atom representation). The analysis of the RMSF plots (Figure \ref{fig: rmsf-all-canvas-at-apo}) shows that, for each cluster, the fluctuations follow the same trend for both all-atom and CANVAS simulations; nonetheless, the RMSD and RMSF values of pembrolizumab in the CANVAS case appear rather low when compared to the ADK simulations, where the multi-resolution model quantitatively reproduces the atomistic fluctuations. This may be ascribed to the interconnections between distant antibody regions, which take place through an extended network of inter-domain correlations within and around the hinge \cite{tarenzi2021communication}. Indeed, we expect that the differences between atomistic and multiscale fluctuations are due to the particularly small high-resolution region chosen for the pembrolizumab with respect to ADK; in the latter case, $\sim 62\%$ of the residues is described atomistically, while in the former case only $\sim 3\%$ of the residues is at high resolution. To test this hypothesis, we performed an additional, 50 ns long CANVAS simulation of pembrolizumab with a larger size of the atomistic region; here, the number of atomistic residues is about the 16\% of the total. The resulting RMSF (Figure S9) shows that including in the high-resolution region also those Fab and Fc residues that are in contact with the hinge leads indeed to a better agreement between the all-atom and CANVAS simulations, with respect to the case where the hinge alone is treated atomistically. In this regard, we stress that the choice of the optimal level and distribution of coarsening to be employed in the construction of a multiple-resolution protein model is a complex and difficult task \textit{per se} \cite{giulini2021system, diggins2018optimal}; nonetheless, CANVAS would represent a powerful instrument to investigate precisely this aspect, in that it allows a simple parametrisation of the model and the subsequent study of the optimal resolution modulation required to correctly and quantitatively reproduce specific system features.

Residue fluctuations are further investigated by computing the linear correlation between the RMSF of C$_\alpha$ atoms of fully-atomistic simulation and the CANVAS one for each case. The latter is given by the calculation of Pearson coefficient\cite{benesty2009pearson} $\rho$, as reported in the scatter plots of Figure S10. All clusters show satisfactory results, with good RMSF correlations ($\rho \sim 0.7$); moreover, an excellent correlation is found in cluster $0_A$ ($\rho \sim 0.87$). In order to gain additional information about the latter result, we also calculated the \textit{cross} Pearson coefficient $\rho_{XY}$ between states and models as summarized in Table \ref{tab: cross-parson-coeff}. $X$ and $Y$ take values in $[0,5]$, corresponding to the various clusters, while each variable is associated to the all-atom ($X$) and CANVAS ($Y$) model. For instance, the value of $\rho_{25}$ corresponds to the Pearson correlation coefficient between the RMSF of C$_\alpha$ atoms for the 2$_A$ cluster at fully atomistic resolution {\it versus} the RMSF of the 5$_A$ state simulated with CANVAS. Diagonal elements $\rho_{XX}$ measure the correlation between C$_\alpha$ atoms of a fully-atomistic simulation and corresponding CANVAS one for the same cluster. Such values, already displayed in the scatter plot of Fig. S5, are highlighted in bold in the table. One can notice that the higher the cluster index, the lower the value of the Pearson correlation coefficient ($\rho_{00} =0.87$, $\rho_{55} =0.68$). Since the clusters are ranked by increasing radius of gyration (or equivalently decreasing compactness), the reason of this trend can be ascribed to the fact that  the CANVAS model of a more open structure has more freedom to explore conformations further and further away from the reference.

Furthermore, Table \ref{tab: cross-parson-coeff} shows that the Pearson coefficient is not systematically higher when comparing simulations starting in the same conformational basin. This is not a fully unexpected result; indeed, CANVAS simulations were started from given initial conformations that in this case are also representative of specific groups of structures sampled in an all-atom MD trajectory, but this gives no guarantee that the whole run will explore the same cluster. This is true in general, even in the case of a fully atomistic model: a new all-atom simulation starting from a representative frame of one conformational cluster might, due to its stochastic nature, diffuse towards another cluster and hence show a fluctuation pattern closer to what is observed in a different set of frames. In the case under examination, additionally, the CANVAS model consists of a distinct structural representation and interaction force field with respect to the all-atom reference; hence, even if the simulation starts from a representative frame of the all-atom cluster, this frame won’t be an equilibrium, representative configuration of the conformational space that would be sampled by the CANVAS model. What we observe in our analysis is that, in spite of these \emph{caveat}, the CANVAS simulations show a remarkable structural overlap between the conformations sampled starting from a given frame and the all-atom cluster they represent, as it can be seen from the CANVAS simulation trajectories provided as SI; as for the pattern of fluctuations, the strong intra-cluster consistency is paired by a non-negligible, and sometimes higher, correlation with different reference clusters, whose appearance is thus not unexpected nor surprising. Hence, while further work is certainly needed to perfect the agreement between the all atom model and its multiple-resolution counterpart, the strong structural consistency and the highly-correlated RMSF patterns of CANVAS runs against their corresponding references support the idea that the model can already capture rather fine details of the molecule's dynamics.

Further analyses were performed to differentiate the dynamics of all-atom and CANVAS simulations for different clusters. Specifically, we have examined the fluctuation correlations distinguishing residues by their level of resolution (AT, MG, CG) and the domain they belong to (FAB1, FAB2, FC). This analysis highlights other salient properties of the fluctuations of the antibody:
%Hence, further analyses are needed to differentiate the dynamics of all-atom and CANVAS simulations for different clusters.

\begin{enumerate}

    \item \textit{Scatter plot with points colored based on resolution} (AT, MG, CG) in Figure S11. The all-atom part is very small ($\sim 3\%$), hence the corresponding value of the Pearson coefficient is not indicative. Conversely, the medium- and coarse-grained parts make up for most of the antibody ($\sim 97\%$), hence the value of $\rho_{\text{MG}}$ and $\rho_{\text{CG}}$ is closer to the one of the full system (dash black line).

    \item \textit{Scatter plot with points colored based on domain partition} (FAB1, FAB2, FC) in Figure S12. Each domain produces a linear pattern in the plot, and the values of the corresponding Pearson coefficients is close to unity. It is worth noticing that, in some of the clusters, the RMSF of the two Fab domains indicates differences in flexibility between the all-atom and the CANVAS models. Specifically, while the overall correlation degree is rather high, the slope of this correlation is different between the two domains. A close inspection reveals indeed that the two heavy chains present a different arrangement of the hinge and of the CH2 domain, as already noted elsewhere~\cite{scapin2015structure, tarenzi2021communication}, thus returning a model whose Fab domains have different interactions and, therefore, different flexibilities.
    
\end{enumerate} 

These analyses provide an additional confirmation that the RMSF correlation between all-atom simulation and the CANVAS one is rather high, although more sophisticated and less straightforward than expected; this, in hindsight, is a reasonable behaviour for a system whose structural and dynamical modules are represented, modelled, and simulated with distinct levels of resolution. 

\begin{table}
\begin{center}
\begin{tabular}{|c|c|c|c|c|c|c|c|}
\cline{3-8}
\multicolumn{2}{c|}{\Large $\rho_{_\text{XY}}$} & \multicolumn{6}{c|}{\textbf{CANVAS}} \\
\cline{3-8}
\multicolumn{2}{c|}{} & $0_A$ & $1_A$ & $2_A$ & $3_A$ & $4_A$ & $5_A$\\
\hline
%\multirow{6}{*}{\begin{sideways} \textbf{all-atom}~ \end{sideways}} & $0_A$ & \textbf{0.87} & 0.58 & 0.50 & 0.27 & 0.63 & 0.35  \\
\multirow{6}{*}{\textbf{all-atom}} & $0_A$ & \textbf{0.87} & 0.58 & 0.50 & 0.27 & 0.63 & 0.35  \\
\cline{2-8}
& $1_A$ & 0.81 & \textbf{0.72} & 0.63 & 0.42 & 0.72 & 0.47 \\
\cline{2-8}
& $2_A$ & 0.73 & 0.60 & \textbf{0.71} & 0.50 & 0.76 & 0.52 \\
\cline{2-8}
& $3_A$ & 0.78 & 0.74 & 0.84 & \textbf{0.74} & 0.91 & 0.74 \\
\cline{2-8}
& $4_A$ & 0.49 & 0.76 & 0.77 & 0.80 & \textbf{0.69} & 0.76 \\
\cline{2-8}
& $5_A$ & 0.55 & 0.83 & 0.64 & 0.66 & 0.61 & \textbf{0.68} \\
\hline
\end{tabular}
\end{center}
\caption{\textit{Cross} Pearson coefficients $\rho_{XY}$ between states and models. $X$ and $Y$ refer to the all-atom and CANVAS model respectively; both indices correspond to the conformation from which the simulations start (0$_A$, 1$_A$, 2$_A$, 3$_A$, 4$_A$, 5$_A$). On the diagonal, the higher the index $XX$, the less compact the antibody conformation, and the lower the value of $\rho_{XX}$.}
\label{tab: cross-parson-coeff}
\end{table}

The conformational dynamics of the system was further inspected by computing the root mean square inner product (RMSIP) between the essential subspaces given by the first $n$ normal modes of the covariance from the atomistic and CANVAS simulations, with $n$ ranging from one to the first 10 modes. A value of 0 indicates that the two mode subspaces are orthogonal, while 1 indicates that they are identical \cite{amadei1999convergence}. Figure S13 shows that less than 5 modes are enough to attain a very good overlap (RMSIP$>$0.8) for all clusters.

We compared the similarity of the structures sampled in the atomistic and CANVAS simulations through the calculation of the radius of gyration (Figure S14). The values present small deviations, with the largest discrepancy of $1.3$ \AA\ observed in cluster 3$_A$; however, in all cases, the radius of gyration from the all-atom simulations is slightly larger than that from the multiscale case, arguably because the steric effects of the side chains cannot be perfectly matched in the very coarse representation employed here, where only the C$_\alpha$ or backbone atoms are retained for more than $97\%$ of the residues.

\begin{figure}[htp]%[htbp]
\centering
\includegraphics[width=\columnwidth]{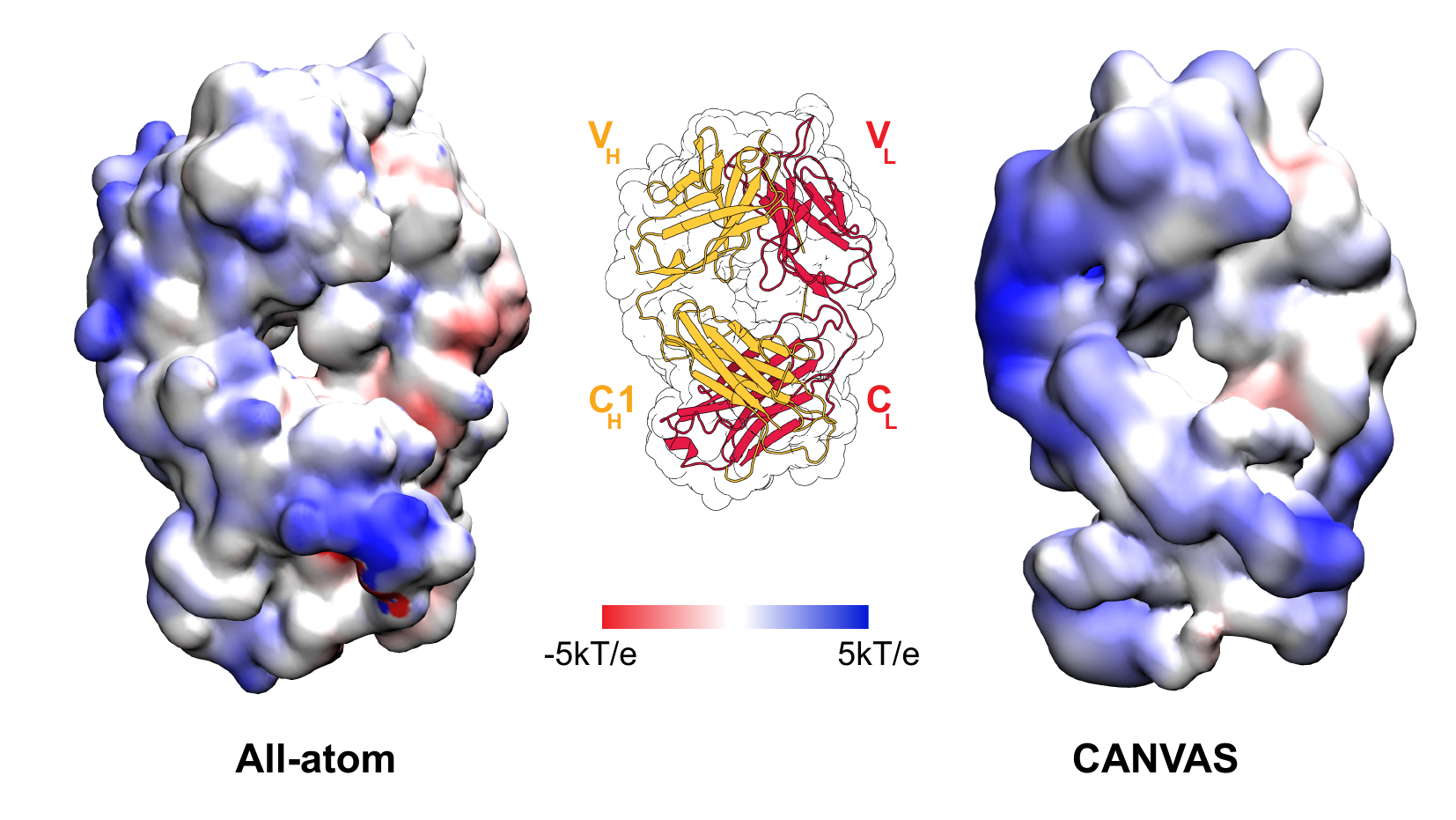}
\caption{Electrostatic potential calculated with the adaptive Poisson-Boltzmann solver (APBS) for the all-atom and CANVAS representations of pembrolizumab Fab1, mapped on the protein surface.}
\label{fig:apbs_antibody}
\end{figure}

As previously done for ADK, the electrostatic potential of the Fab1 domain at MG/CG resolution has been computed for the antibody Fab, on the basis of the partial charges assigned to each bead in the CANVAS model (Figure \ref{fig:apbs_antibody}). The comparison between the all-atom and low-resolution case shows a good similarity.

\begin{figure*}[htp]%[htbp]
\centering
\includegraphics[width=\textwidth]{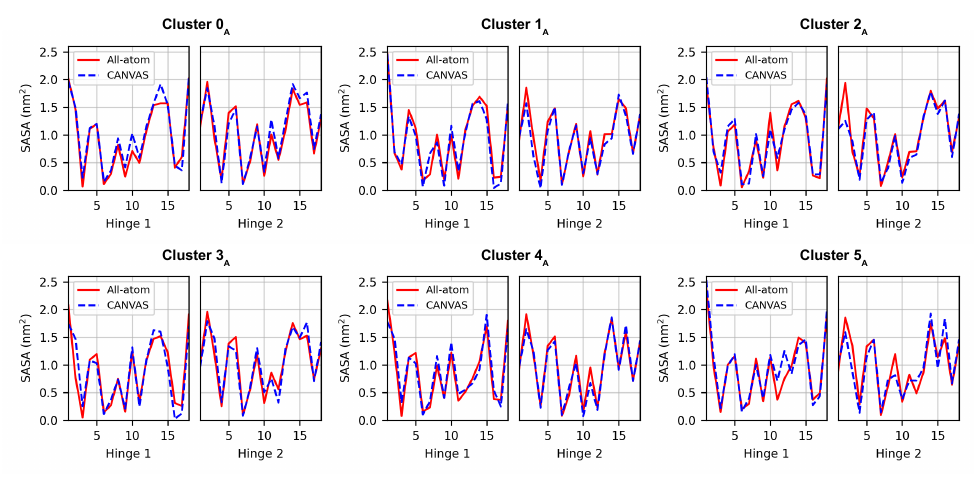}
\caption{Per-residue values of SASA, computed for the atomistic region of the antibody in each conformational basin. Hinge 1 belongs to chain B, while hinge 2 belongs to chain D.}
\label{fig:sasa-hinge}
\end{figure*}

The average SASA was computed along the trajectory for each residue of the atomistic region, namely the two hinge segments (Figure \ref{fig:sasa-hinge}). The comparison between the SASA values computed from the all-atom and multiscale simulations, performed for each conformational cluster, shows a very good agreement. The CANVAS model proves able to accurately reproduce the solvent exposure of the atomistic residues in relation to the conformational properties of the fully atomistic system.

\section{Conclusions}\label{sec:conclusions}

In this work we introduced the CANVAS model for the multiscale molecular dynamics simulation of proteins. The model couples different levels of detail within the same protein representation, ranging from a fully atomistic description to a coarse one, e.g. one bead per amino acid (as in the case studies discussed here) or even lower levels of resolution. CANVAS allows a smooth transition between these resolutions, by including regions at intermediate levels of detail. Importantly, the non-bonded components of the interaction potential are taken into account at all resolution levels, by assigning to each CG bead the average properties -- including charge, size, and dispersion energy -- of the atoms that are mapped onto it. This property enables, in principle, the application of CANVAS for the simulation of large, multimeric protein complexes, where also the CG resolution can be used to model realistic molecular interfaces. This application will be explored in future works.

Here, we have tested the CANVAS model on two systems of very different size and conformational dynamics, namely the enzyme adenylate kinase and the therapeutic antibody pembrolizumab. To validate the model, we performed a comparison between properties extracted from the fully atomistic and the multiscale simulations, in terms of residue fluctuations, large-scale dynamics, solvent exposure, and electrostatic properties; in all cases, the CANVAS model results are in good agreement with the all-atom reference.

The variable-resolution modelling approach presented here achieves two key goals: first, it demonstrates that a sensible modulation of the resolution can be employed to construct models of large molecules whose behaviour is the same of, or quantitatively consistent with, that of a reference all-atom model of the system; second, it enables the rapid, practical construction of tailored low-resolution models of such molecules with minimal information and no reference simulations. The possible applications of these models cover a broad spectrum; we here stress those that appear most promising to us, namely the exploration of the conformational space of molecules whose structure is known with low resolution only, or the characterisation of the structure-dynamics-function relation by means of the systematic modulation of the resolution throughout the structure. An additional future application is the efficient calculation of binding free energies, employing an atomistic accuracy only in the active and/or allosteric sites. The relevance of taking into account distant protein domains within the simulation setup has been proven in various cases, as e.g. specifically observed in the case of antigen-antibody binding affinities~\cite{knapp2017variable, tarenzi2021communication}; the possibility of keeping a simplified description of the vast majority of the molecule thus represents an advantage with respect to the alternative approach of simulating only the protein domain involved in the binding. In addition, we can expect that the impact on entropy due to the reduction of the number of degrees of freedom is similar, and therefore does not affect the result, if the aim is to compute relative binding free energy among a set of similar ligands, where the mapping of the protein is kept the same. All the abovementioned applications, which involve the usage of the CANVAS model in combination with efficient simulation methods for the solvent (e.g. multi-timestepping \cite{humphreys1994multiple, krieger2015new}, implicit solvent \cite{onufriev2019generalized, chen2021machine}, or adaptive resolution simulation methods \cite{praprotnik2007adaptive, Donadio_PRL_2013-hadres_molliq, tarenzi2017open}) are currently under development, and pave the way to a novel approach to computer-aided molecular biochemistry.


\section*{Data and software availability}
The CANVAS software is available for download at \href{https://github.com/potestiolab/canvas}{https://github.com/potestiolab/canvas}, including the manual and tutorials. The raw data produced and analysed in this work are freely available on the Zenodo repository \href{https://doi.org/10.5281/zenodo.7225086}{https://doi.org/10.5281/zenodo.7225086}. 

\section*{Acknowledgement}
The authors are indebted with Roberto Menichetti for the calculation of the distance-dependent elastic constants, and with Giovanni Mattiotti for an insightful reading of the manuscript. This project received funding from the European Research Council (ERC) under the European Union's Horizon 2020 research and innovation program (Grant 758588).

\section*{Author contributions}
RP designed the project; RF developed the software and ran the simulations; RF and TT performed the analyses. All authors contributed to the interpretation of the results and to the writing of the manuscript.

%\section*{Terms \& Conditions}
%Most electronic Supporting Information files are available without a subscription to ACS Web Editions. Such files may be downloaded by article for research use (if there is a public use license linked to the relevant article, that license may permit other uses). Permission may be obtained from ACS for other uses through requests via the RightsLink permission system: http://pubs.acs.org/page/copyright/permissions.html.

\section*{Notes}
The authors declare no competing financial interest.

\section*{Supporting information}

%A listing of the contents of each file supplied as Supporting Information
%should be included. For instructions on what should be included in the
%Supporting Information as well as how to prepare this material for
%publications, refer to the journal's Instructions for Authors.

The Supporting Information is available free of charge.
\begin{itemize}
  \item Derivation of the elastic constants $k_{nb}$.
  \item Supplementary figures: RMSD and RMSF of ADK using GROMACS/LAMMPS and Charmm/Amber force fields; RMSF from atomistic simulations of ADK using reaction-field or PME; distance between C$_\alpha$ atoms of residues A55 and V169 in the all-atom simulation of ADK; alignment of atomistic and CANVAS representative structures, in open and closed conformations; RMSF of ADK from atomistic and CANVAS simulations, in open and closed conformations; representative structures of pembrolizumab antibody; RMSD from pembrolizumab simulations; RMSF from pembolizumab simulations with different sizes of the atomistic region; scatter plots of RMSF for each pembrolizumab cluster, and calculation of Pearson coefficient; scatter plots colored on the basis of the resolution and of the structural domain; RMSIP between the essential subspaces computed from the atomistic and CANVAS simulations; average radius of gyration for each conformational cluster of the antibody.
\end{itemize}


\bibliography{main_v2}

%%%%%%%%%%%%%%%%%%%%%%%%%%%%%%%%%%%%%%%%%%%%%%%%%%%%

\end{document}